\documentclass[runningheads,fleqn]{svmult}

\usepackage{makeidx}   
\usepackage{graphicx}  
\usepackage{subeqnar}  
\usepackage{multicol}  
\usepackage{physproc}  
\makeindex             

\def\r{{\bf{r}}}

\def\k{{\bf{k}}}
\def\kt{\tilde{\k}}
\def\K{{\bf{K}}}

\def\M{{\bf{M}}}

\def\tk{\tilde{\bf{k}}}

\def\tq{\tilde{\bf{q}}}

\def\ep{\epsilon}

\def\ep{\epsilon}

\begin{document}
\title*{The Dynamical Cluster Approximation: \protect\newline A New
Technique for Simulations of \protect\newline
 Strongly Correlated Electron Systems}
\toctitle{The Dynamical Cluster Approximation } 
%
%
\titlerunning{The Dynamical Cluster Approximation}
%
\author{S. Moukouri
\and C. Huscroft
\and M. Jarrell}
\authorrunning{S. Moukouri et al.}
%
%
\institute{University of Cincinnati, Cincinnati OH 45221, USA}

\maketitle              

\begin{abstract}
We present the algorithmic details of the dynamical cluster approximation
(DCA) algorithm.  The DCA is a fully-causal approach which systematically
restores non-local correlations to the dynamical mean field approximation
(DMFA).  The DCA is in the thermodynamic limit and becomes exact for
an infinite cluster size, while reducing to the DMFA for a cluster size of
unity.  Using the one-dimensional Hubbard Model as a non-trivial test
of the method, we systematically compare the results of a quantum Monte
Carlo (QMC) based DCA with those obtained by finite-size QMC simulations
(FSS).  We find that the single-particle Green function and the self-energy
of the DCA and FSS approach the same limit as the system size is 
increased, but from complimentary directions.  The utility of the DCA in
addressing problems that have not been resolved by FSS is demonstrated.

\end{abstract}

\section {Introduction}

One of the most active subfields in condensed matter theory is the 
development of new algorithms to simulate the many-body problem. This 
interest is motivated by various physical phenomena, including high 
temperature superconductivity, magnetism, heavy fermions and the rich 
phenomenology occurring in quasi-one dimensional compounds. In the last 
few years, important progress has been made. Well-controlled results 
have been obtained by exact diagonalization and quantum Monte Carlo 
methods (QMC)\cite{dagotto}.  However, these algorithms suffer from a 
common limitation in that the number of degrees of freedom grows rapidly 
with the lattice size.  As a consequence, the calculations are restricted 
to relatively small systems. In most cases, the limited size of the 
system prohibits the study of the low-energy physics of these models.

Recently, another route to quantum simulations has been proposed. 
Following Metzner and Vollhardt\cite{metzvoll} and 
M\"uller-Hartmann\cite{muller-hartmann} who showed  that in the limit 
of infinite dimensions, the many-body problem becomes purely local, 
a mapping to a self-consistent Anderson impurity problem was 
performed \cite{pruschke}\cite{georges}. The availability of many 
techniques to solve the Anderson impurity Hamiltonian has led to a 
dramatic burst activity.   However, when applied to systems in two 
or three dimensions this self-consistent approximation, referred to 
as the dynamical mean field approximation (DMFA), displays some
limitations. Due to its local nature, the DMFA neglects spatial
fluctuations which are essential when the order parameter is non-local. 
This is the case in the cuprate or heavy fermion superconductors. Methods 
that at least include short-range spatial fluctuations are currently 
the object of intensive research. These techniques map the lattice
problem to a self-consistently embedded finite-sized cluster, instead 
of a single impurity as in DMFA.

The most promising of these techniques is the dynamical cluster
approximation (DCA) \cite{DCA_hettler}\cite{DCA_hettler2}\cite{DCA_maier1}. 
The key idea of the DCA is to use the irreducible quantities (self energy, 
irreducible vertices) of the embedded cluster as an approximation 
for the lattice quantities. These irreducible quantities are then applied 
to construct the lattice reducible quantities such as the Green function 
or susceptibilities in the different channels. The cluster problem is 
solved by using a generalization of the Hirsch-Fye quantum Monte Carlo 
(QMC) method.

In this paper, we discuss the novelty and the new perspectives offered
by the DCA.  We use the one-dimensional Hubbard model as a test
case. We analyze the convergence of the DCA in comparison to that of
the usual finite systems simulations (FSS). The choice of a one-dimensional
model is particularly instructive. A success of the DCA in 1D, where quantum
fluctuations are the strongest, means that the method can work in any
dimension.  We show that the DCA produces results which are complimentary
to those obtained by FSS.

\begin{figure}[b]
\includegraphics[width=.6\textwidth]{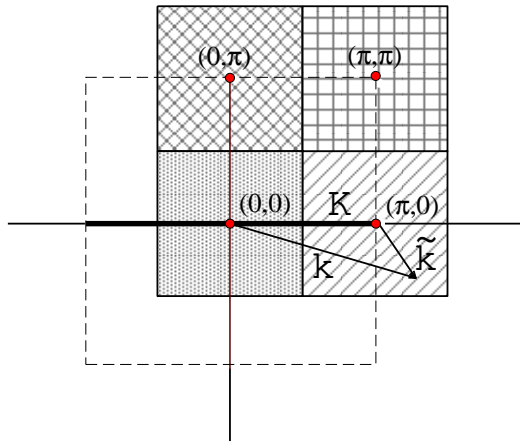}
\caption{$N_c=4$ cluster cells (shown by different fill patterns) that
partition the first Brillouin Zone (dashed line).  Each cell is centered
on a cluster momentum $\K$ (filled circles). To construct the DCA cluster, we
map a generic momentum in the zone such as $\k$ to the nearest cluster point
$\K=\M(\k)$ so that $\kt=\k-\K$ remains in the cell around $\K$.  }
\label{BZ}
\end{figure}

\section{The Dynamical Cluster Approximation}

\subsection{Formalism}

The DCA is based on the assumption that the lattice self energy is weakly
momentum dependent.  This is equivalent to assuming that the dynamical
intersite correlations have a short spatial range $ < L/2$ where $L$ is
the linear dimension of the cluster.  Then, according to Nyquist's sampling
theorem\cite{nyquist}, to reproduce these correlations in the self energy,
we only need to sample the reciprocal space at intervals of
$\Delta k\approx 2\pi/L$.  Therefore, we could approximate $G(\K+\kt)$ by
$G(\K)$ within the cell of size $\left( \pi/L\right)^D$ (see, Fig.~\ref{BZ})
centered on the cluster momentum $\K$ (wherever feasible, we suppress the
frequency labels) and use this Green function to calculate the self energy.
Knowledge of these Green functions on a finer scale in momentum is unnecessary,
and may be discarded to reduce
the complexity of the problem. Thus the cluster self energy can be constructed
from the {\em{coarse-grained average}} of the single-particle Green function
within the cell centered on the cluster momenta:
\begin{equation}
\bar{G}(\K) \equiv \frac{N_c}{N}\sum_{\kt}G(\K+\kt) \quad\mbox{,} 
\label{gbar}
\end{equation}
where $N$ is the number of points of the lattice, $N_c$ is the number of
cells in the cluster, and the $\kt$ summation runs over the momenta of
the cell about the cluster momentum  $\K$ (see, Fig.~\ref{BZ}).  For short
distances $r< L/2$ the Fourier transform of the Green function
$\bar{G}(r) \approx G(r) +{\cal{O}}((r\Delta k)^2)$, so that short ranged
correlations are reflected in the irreducible quantities constructed from
$\bar{G}$; whereas, longer ranged correlations $r>L/2$ are cut off by the
finite size of the cluster.\cite{DCA_hettler}

\begin{figure}[b]
\includegraphics[width=.8\textwidth]{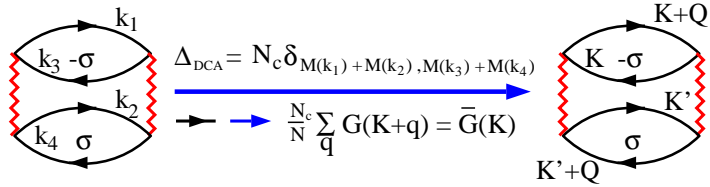}
\caption{  The DCA choice of the Laue function Eq.~\ref{lauedca} leads to
the replacement of the lattice propagators $G({\bf k}_1)$,
$G({\bf k}_2)$,... by coarse grained propagators $\bar{G}({\bf K})$,
$\bar{G}({\bf K}^\prime)$, ... (Eq.~\ref{gbar}) in the internal
legs of $\Phi_{DCA}$, illustrated for a second order diagram.}
\label{collapse_DCA}
\end{figure}

This coarse graining procedure and the relationship of the DCA to the DMFA
is illustrated by a microscopic diagrammatic derivation of the DCA.  For
Hubbard-like models, the properties of the bare vertex are completely
characterized by the Laue function $\Delta$ which expresses the momentum
conservation at each vertex.  In a conventional diagrammatic approach
\begin{eqnarray}
\Delta(\k_1,\k_2,\k_3,\k_4)=
\sum_\r \exp\left[i\r\cdot(\k_1-\k_2+\k_3-\k_4)\right]=
N \delta_{\k_1+\k_2,\k_3+\k_4}
\end{eqnarray}
 where $\k_1$ and $\k_2$ ($\k_3$ and
$\k_4$) are the momenta entering (leaving) each vertex through its
legs of $G$.  However as $D\to\infty$ M\"uller-Hartmann showed that the
Laue function reduces to\cite{muller-hartmann}
\begin{eqnarray}
\Delta_{D\rightarrow\infty}({\bf k}_1,{\bf k}_2,{\bf k}_3,{\bf k}_4)=
1+{\cal O}(1/D)\quad\mbox{.}
\label{Laueinft}
\end{eqnarray}
The DMFA assumes the same Laue function,
$\Delta_{DMFA}(\k_1,\k_2,\k_3,\k_4)=1$,
even in the context of finite dimensions.  Thus, the conservation of momentum
at internal vertices is neglected.
Therefore we may freely sum over the internal momenta at each vertex in the
generating functional $\Phi_{DMFA}$. This leads to
a collapse of the momentum dependent contributions to the functional
$\Phi_{DMFA}$ and only local terms remain.

The DCA systematically restores the momentum conservation at internal 
vertices.  The Brillouin-zone is divided into $N_c=L^D$ cells of size 
$(2\pi/L)^D$.
Each cell is represented by a cluster momentum $\bf K$ in the center of
the cell. We require that momentum conservation is (partially) observed
for momentum transfers between cells, i.e., for momentum transfers larger
than $\Delta k=2\pi/L$, but neglected for momentum transfers within a
cell, i.e., less than $\Delta k$. This requirement can be established by
using the Laue function \cite{DCA_hettler}
\begin{equation}
\Delta_{DCA}(\k_1,\k_2,\k_3,\k_4)=
N_c \delta_{\M(\k_1)+\M(\k_3),\M(\k_2)+\M(\k_4)} \quad\mbox{,}
\label{lauedca}
\end{equation}
where $\M(\k)$ is a function which maps $\k$ onto the momentum label $\K$
of the cell containing $\k$ (see, Fig.~\ref{BZ}).  
With this choice of the Laue function the momenta of each internal leg may
be freely summed over the cell.  This is illustrated for the second-order
term in the generating functional in Fig.~\ref{collapse_DCA}.  Thus, each
internal leg $G(\k_1)$ in a diagram is replaced by the coarse--grained Green
function ${\bar G}(\M(\k_1))$, defined by Eq.~\ref{gbar}.
The diagrammatic sequences for the generating functional and its
derivatives are unchanged; however, the complexity of the problem is
greatly reduced since $N_c\ll N$.  We showed
previously\cite{DCA_hettler,DCA_maier1} that the DCA estimate of
the lattice free-energy is minimized by the approximation
$\Sigma(\k)\approx {\bar{\Sigma}}({\bf M}({\bf k}))+{\cal O}(\Delta k^2)$,
where $\delta\Phi_{DCA}/\delta\bar{G}=\bar{\Sigma}$.

One advantage of the DCA over other attempts to build 
self-consistent cluster techniques is that the DCA is fully causal.
The spectral weight is conserved and the imaginary part of the 
single-particle retarded Green function and self-energy are negative
definite. Hettler et al. used a geometric argument to derive
a rigorous proof of the causality of the DCA \cite{DCA_hettler}.
 
\subsection{One and two-particle quantities}

\begin{figure}[b]
\includegraphics[width=.58\textwidth]{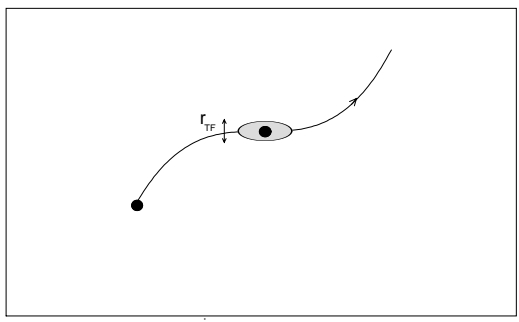}
\caption{  Motion of a particle with its screening cloud}
\label{lendca}
\end{figure}

In the DMFA, after convergence, the local Green function of
the lattice is identical to that of the impurity model. Though in the
DCA, the coarse-grained Green function ${\bar G(K)}$ is equal to
the cluster Green function, this quantity is not used as an
approximation to the true lattice Green function $G(K)$.  The correct 
procedure to calculate the lattice physical quantities within the DCA 
is to approximate the lattice irreducible quantities with those of the 
cluster. The lattice reducible quantities are then deduced from the 
irreducible.  In order to completely understand the DCA formalism, one
must understand why reducible and irreducible quantities are treated
differently.  Consider a screened particle (a quasiparticle) propagating
through the system, as pictured in Fig.~\ref{lendca}.  Here, we will assume
that the screening length, e.g., due to Thomas-Fermi screening, $r_{TF}$ is
short.  This screening cloud is described by the single-particle
self energy $\Sigma(\k,\omega)$ which itself may be considered
a functional of the interaction strength $U$ and the single-particle
propagator $G(\k,\omega)$.  The different screening processes are
described perturbatively by a sum of self energy diagrams.  Due to the
short screening length, the propagators which describe these processes
need only be accurate for lengths $< r_{TF}$.  From the Fourier
uncertainty principle, we know that the propagators at short lengths
may be accurately described by a coarse sampling of the reciprocal
space, with sampling rate $\Delta k= \pi/r_{TF}$.  On the other
hand, the phase accumulated as the particle propagates through 
the system is described by the Fourier transform of the single-particle 
Green function.  Since this accumulated phase is crucial in the 
description of the quantum dynamics it is important that $G(r)$ 
remain accurate at long length scales, so it should not be 
coarse-grained as described above.   However
it may be constructed from the approximate self energy.
Hence, the lattice Green function is given by,
\begin{eqnarray}
G({\bf k})=\frac{1}{i\omega_n - \ep_{{\bf k}}-\Sigma( M({\bf k}))}\quad\mbox{.}
\label{latg}
\end{eqnarray} 

A similar procedure is used to construct the two-particle quantities 
needed to determine the phase diagram or the nature of the dominant
fluctuations that can eventually destroy the single-particle quasi-particle.
They are computed in the same fashion as the single-particle ones. The 
irreducible vertex $\Gamma$ is the analogue of the self energy for 
two-particle quantities. In a general many-body theory, $\Gamma$ has 
four entries describing the states of a pair before and after the
interaction,
\begin{eqnarray}
\Gamma=\Gamma(k_1,k_2,k_3,k_4) \quad\mbox{.}
\label{gamma}
\end{eqnarray}
  The diagonalization of $\Gamma$ in spin space reduces
to the usual density and magnetic fluctuations in the particle-hole channel
or singlet and triplet fluctuation in the particle-particle channel. 
For the sake of simplicity, we will use a generic notation below which
describes any one of these four fluctuations.
Momentum conservation ensures that there are only three independent momenta, 
$\Gamma=\Gamma(q,k,k')$.  We assume, as for the self energy, that the 
irreducible vertex $\Gamma= \frac {\delta \Sigma}{\delta \bar {G}}$ is 
weakly momentum dependent. Hence, this quantity is equated to the 
corresponding cluster quantity within a cell surrounding the cluster 
momentum for each momentum involved in $\Gamma(q,k,k')$. This gives,
\begin{eqnarray}
\Gamma_{\bf Q+\tq}({\bf K+\tk},{\bf K'+\tk'})= 
\Gamma_{\bf Q}({\bf K},{\bf K'})  \quad\mbox{.} 
\label{gamclus}
\end{eqnarray}

The calculation procedure of the reducible two-particle quantities from 
the irreducible vertex $\Gamma$ will now be described.  The QMC cluster 
susceptibility 
\begin{eqnarray}
 \chi_c(Q,K,K')=\langle c_{K+Q}^{\dag}(\tau)c_{K'-Q}(\tau')c_K^{\dag}(0)c_{K'}(0) \rangle  \quad\mbox{,} 
\label{chiclus}
\end{eqnarray}
calculated in the appropriate channel is the analogue of the single-particle
Green function $G_c$. But this quantity is different from the coarse grained 
susceptibility $\bar \chi$,
\begin{eqnarray}
\bar {\chi}_{\bf Q+\tq}({\bf K},{\bf K'}) = \frac {N_{c}^{2}} {N^2} 
\sum_{\bf \tk,\tk'}
 \chi_{\bf Q+ \tq}({\bf K+\tk},{\bf K'+ \tk'})  \quad\mbox{,} 
\label{chicoars}
\end{eqnarray}
 where $\chi$ is the lattice susceptibility.  In the DCA, ${\bar \chi}$ and 
$\chi_c$ are both related to $\Gamma$ through the reduced Bethe-Salpeter 
equation,
\begin{eqnarray}
{\chi}_{\bf Q+\tq}^{\alpha}= \chi_{\bf Q+\tq}^{0\alpha}
 + \chi_{\bf Q+\tq}^{0\alpha} \Gamma_Q 
 {\chi}_{\bf Q+\tq}^{\alpha}  \quad\mbox{,} 
\label{bethe}
\end{eqnarray}
$\chi^{\alpha}$ represents either $\chi_c$ or $\bar {\chi}$.
$\bar {\chi}_{\bf Q+\tq}^0$ is obtained from the single-particle
Green function as follows,
\begin{eqnarray}
\bar {\chi}_{\bf Q+\tq}^0 =  \delta_{{\bf K,K'}}\frac{N_c^2}{N}
\sum_{\tk}G({\bf K+\tk})G({\bf K+\tk+Q+\tq})  \quad\mbox{.} 
\label{chinot}
\end{eqnarray}
In Eq.~\ref{bethe} above, the sum is restricted to cluster momenta. 
This constitutes a significant simplification of the lattice problem which 
may be otherwise intractable. Yet, as the cluster size increases, the full 
solution of this reduced Bethe-Salpeter equation may require a significant 
amount of computer storage and CPU time. We thus restrict ourselves to 
momenta $\bf Q$ where a given type of fluctuation is likely to diverge, i.e., 
${\bf Q}=0$ for the density, singlet or ${\bf Q}=(\pi, \pi)$ magnetic
fluctuations.  The inversion of each of the two equations then yields
\begin{eqnarray}
\bar {\chi}^{-1}= \chi_c^{-1}-\chi_c^{0-1}+ \bar {\chi}^{0-1}  \quad\mbox{.} 
\label{chicoars2}
\end{eqnarray}
The charge $(ch)$ and spin $(sp)$ susceptibilities $\tilde {\chi}_{ch,sp}$
are deduced from $\bar {\chi}$
\begin{eqnarray}
\tilde {\chi}_{ch,sp}({\bf q},i\omega_n)=\frac {(k_BT)^2}{N_c^2}
\sum_{{\bf KK'}mm'\sigma\sigma'}
\lambda_{\sigma \sigma'}\bar {\chi}_{\bf q}({\bf K},i\omega_m;{\bf K'},i\omega_{m'})
  \quad\mbox{,} 
\label{chichsp}
\end{eqnarray}
where $\lambda_{\sigma \sigma'}=1$ for the charge channel and 
$\lambda_{\sigma \sigma'}=\sigma\sigma'$ for the spin channel.


\subsection{The DCA algorithm}
\begin{table}
\caption{Steps of the DCA algorithm}

{\bf 0.} set $\Sigma = 0$ 

{\bf 1.} calculate the coarse-grained Green function $\bar G$

{\bf 2.} calculate ${\cal G}^{-1}={\bar G}^{-1} + \Sigma$

{\bf 3.} using Hirsch-Fye calculate the cluster Green function $G_c$

{\bf 4.} compute a new estimate of $\Sigma= {\cal G}^{-1}-G_c^{-1}$

{\bf 5.} repeat {\bf 1-4} until $\Sigma$ converges 

{\bf 6.} accumulate bins of measurements on $G_c$, ${\chi_c}$ 

{\bf 7.} calculate the lattice Green function $G$, spectral weight,
susceptibilities...
\label{tab}
\end{table}

The DCA iteration procedure is set forth in Table~\ref{tab}. It is started 
by setting the initial self energy to zero. This self energy is then used 
to compute  the coarse-grained Green function $\bar {G}(\bf K)$. The latter 
is used to compute the host Green function
 ${\cal{G}}(\K)^{-1}=\bar{G}(\K)^{-1}+
\Sigma(\K) $ which must be introduced to avoid over-counting
diagrams.  ${\cal {G}}$ serves as the input to the QMC simulation to yield 
a new estimate of the cluster self energy. The procedure is repeated until 
$\Sigma$ converges. This typically happens in less than ten iterations. The 
number of iterations decreases when $N_c$ increases since the coupling
to the host is smaller for larger clusters ($\Gamma \approx {\cal O}(1/N_c)$)
\cite{DCA_maier1}. 
The convergence test is made on the ratio $\rho$,
\begin{eqnarray}
\rho=\frac{|\sum_n (\Sigma_{new}(i\omega_n)-\Sigma_{old}(i\omega_n))|}
{|\sum_n \Sigma_{old}(i\omega_n)|}  \quad\mbox{.} 
\label{rho}
\end{eqnarray}    
\noindent Once convergence is reached, the remaining single and 
two-particle measurements are made in a final QMC iteration. As in a usual 
QMC simulation, bins of measurements are accumulated as discussed in the 
next section. One should, however, bear in mind that these measurements 
are performed with ${\bar G}$ instead of the true lattice Green function 
$G$. Hence, the determination of the lattice quantities requires an 
additional step in which the coarse-graining equation is inverted.  This 
is done in a separate program.

\section {The Hirsch-Fye QMC algorithm}

The cluster problem is solved using one of the available finite system 
techniques.  The Hirsch-Fye (HF) algorithm \cite{fye}, which was originally 
proposed for impurity problems, is easily adapted to embedded cluster 
simulations as required in the DCA.  We recall here the main steps of its 
derivation. For details we refer the reader to the original Hirsch and Fye 
paper or the review by Georges  {\it et al} \cite{georges}.  One starts by 
writing the partition function in the Grand-canonical ensemble,
\begin{eqnarray}
Z=Tr{e^{[-\beta (H-\mu N)]}}  \quad\mbox{.} 
\end{eqnarray}  
Then the exponential is broken up into $N_{\tau}$ imaginary-time slices
and the Trotter approximation is employed to separate the kinetic, K, and 
interacting, V, parts of the Hamiltonian H,
\begin{eqnarray}
e^{[-\beta (H-\mu N)]} = \prod_{l=1}^{N_{\tau}} e^{[-\Delta \tau
(K_l+V_l)]} \approx
\prod_{l=1}^{N_{\tau}} e^{-\Delta \tau K_l} e^{-\Delta \tau V_l}  \quad\mbox{,}  
\end{eqnarray} 
where $N_{\tau} \Delta_{\tau}=\beta$.
The error made during the last step is of the order of $(\Delta \tau)^2$. 
The next step is to replace this interacting problem by one consisting 
of non-interacting particles moving in a fluctuating field $x_{\alpha}$, 
where the index $\alpha$ stands for space and imaginary-time
coordinates. This is done by employing the Hirsch-Hubbard-Stratonovitch 
(HHS) transformation,
\begin{eqnarray}
e^{[-\Delta_{\tau}U(n_{\alpha \uparrow}-\frac{1}{2})
(n_{\alpha \downarrow}-\frac{1}{2})]} = 
\frac {e^{[-\Delta_{\tau}\frac{U}{4}]}}{2}
\sum_{x_{\alpha}=\pm 1} e^{[ \lambda x_{\alpha}(n_{\alpha \uparrow} - n_{\alpha \downarrow})]}  \quad\mbox{.} 
\end{eqnarray}
After the HHS transformation, the quartic operator $V$ becomes quadratic like 
$K$. In this new form a trace over the fermion degrees of freedom can be 
exactly performed. It yields,
\begin{eqnarray}
Z=\sum_{{x_{\alpha}}} \prod_{\sigma=\uparrow \downarrow}
det[(G_c^{\sigma}({x_{\alpha
}}))^{-1}]  \quad\mbox{,} 
\end{eqnarray}          
where the sum is over the configurations of HHS fields ${x_{\alpha}}$. 
The dimension of the Green function matrix 
$G_c^{\sigma}$ is $(N_{\tau} \times N_c)^2$.
It is at this point that the HF algorithm differs from the 
Blanckenbecler-Sugar-Scalapino (BSS) algorithm\cite{bss}. 
In the HF, the simulation deals directly with the matrix $G_c^{\sigma}$. In 
the BSS the locality of the action in time is used to reduce the
matrix-size in the simulation.  However, this cannot be done when
action is non-local in time, as in this case where the cluster is
coupled to a host. 

The algorithm starts with an initial Green function $G_c={\cal{G}}$
and a corresponding initial configuration of the fields ${x_{\alpha}}$.
It then sweeps the space-time lattice by proposing flips in the fields
${x_{\alpha}}\to {x'_{\alpha}}=-{x_{\alpha}}$.  The heat-bath or the 
Metropolis algorithm is used to determine if the change will be accepted.
The probability $P({x_{\alpha}})$ of the configuration ${x_{\alpha}}$ is 
proportional to
$\prod_{\sigma=\uparrow \downarrow}det[(G_c^{\sigma}({x_{\alpha
}}))^{-1}]$.  The transition probability is
\begin{eqnarray}
R= \frac { \prod_{\sigma=\uparrow \downarrow}det[({G'}_{c}^{\sigma}({x_{\alpha
}}))^{-1}]}{ \prod_{\sigma=\uparrow \downarrow}det[(G_{c}^{\sigma}({x_{\alpha
}}))^{-1}]}  \quad\mbox{.} 
\end{eqnarray}         
If the new configuration is accepted, the Green function is updated 
by using the relation
\begin{eqnarray}
{G'}^{\sigma}_{c {\ i,j}}=G^{\sigma}_{c \ {i,j}}
+ \frac {(G^{\sigma}_{c \ {j,k}}-\delta_{i,k})
e^{[-\lambda \sigma (V-V')]}} {1+(1-G^{\sigma}_{c \ {k,k}})
(e^{-\lambda \sigma (V-V')}-1)}G^{\sigma}_{c \ {k,j}}  \quad\mbox{.} 
\end{eqnarray}

The initial field configuration is chosen with all ${x_{\alpha}}=0$. The
above equation is then used to construct a physically realistic field
configuration (i.e., ${x_{\alpha}}=1$ for all fields). The system is then 
warmed up by sequentially stepping through the space-time lattice, proposing
changes at each space-time site $x_{\alpha} \rightarrow -x_{\alpha}$. The 
change is accepted if the transition probability is greater than a random number
between 0 and 1. Typically, the warm up phase lasts for about a  hundred space-time
sweeps before measurements begin. It is necessary to perform a few complete
space-time sweeps in order to produce more-or-less independent measurements. 
For clusters, the Hirsch-Fye algorithm is very efficient and stable at
low temperatures. 

\begin{figure}[b]
\includegraphics[width=\textwidth]{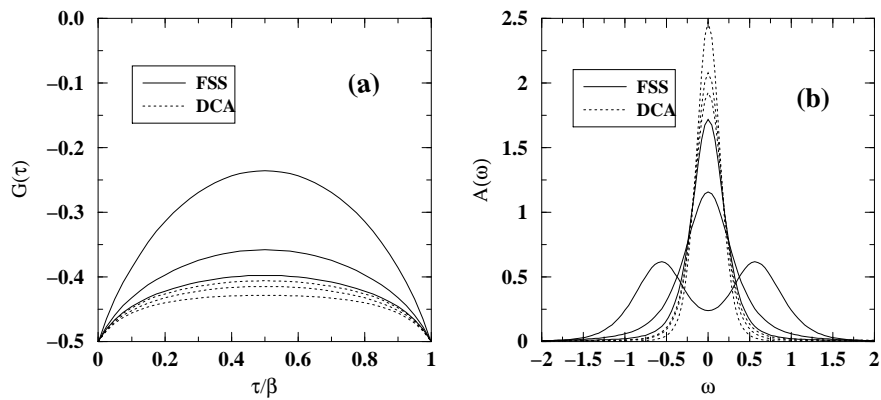}
\caption{ (a) Imaginary-time Green function at the Fermi point for 
$U=2$, $\beta=5$ for FSS
(solid lines) for $N_c$=4,8,16 (from top to bottom) and for DCA
 (dotted lines) for $N_c$=4,8,12 (from bottom to top). (b) The corresponding 
spectral weights for FSS 
(solid lines) $N_c$=4,8,12 (increasing value at $\omega = 0$) 
and DCA (dotted lines) $N_c$=4,8,12 (decreasing value at $\omega = 0$).}
\label{gwu2b5}
\end{figure}
\begin{figure}[b]
\includegraphics[width=\textwidth]{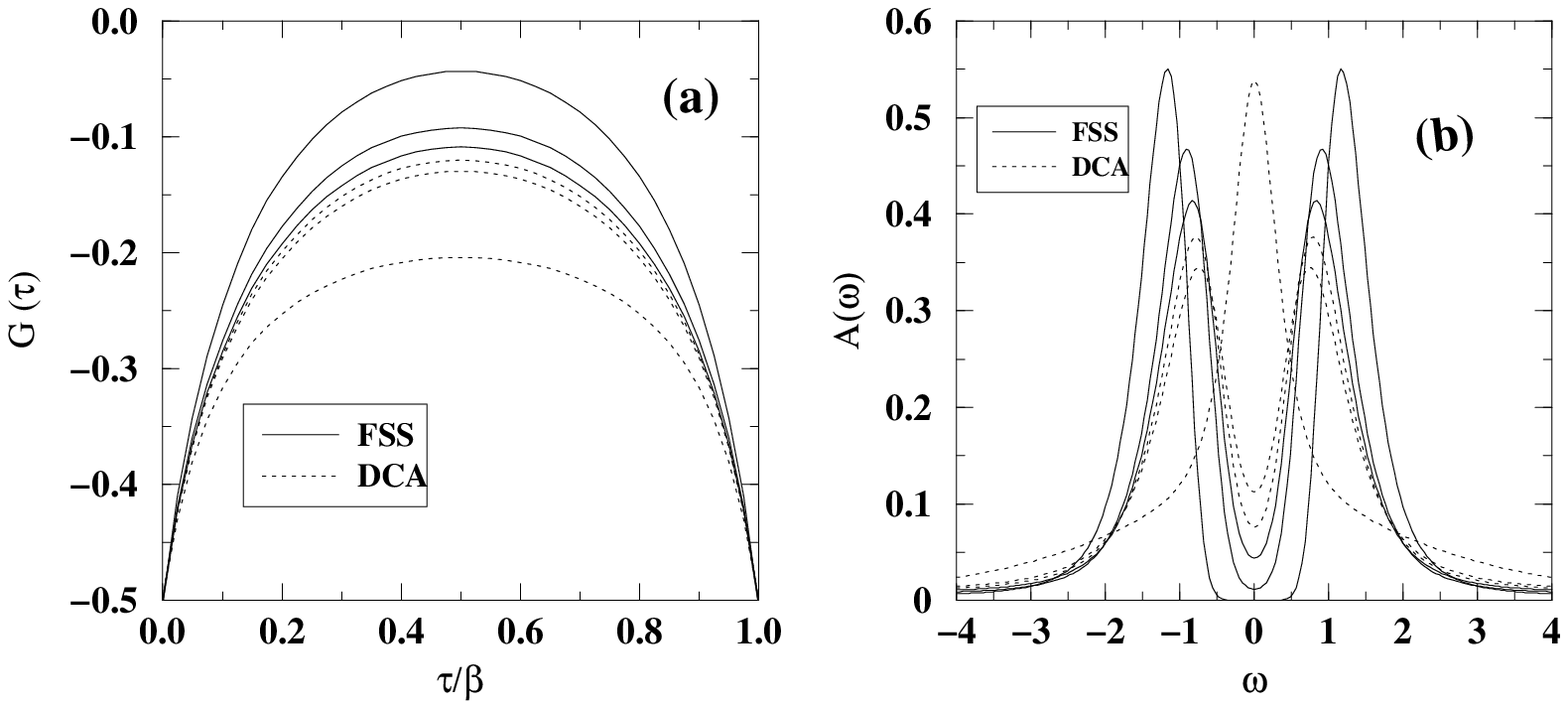}
\caption{  (a) Imaginary-time Green function at the Fermi point for 
$U=4$, $\beta=5$ for FSS
(solid lines) for $N_c$=4,8,16 (from top to bottom) and for DCA
 (dotted lines) for $N_c$=4,8,12 (from bottom to top). (b) The corresponding 
spectral weights for FSS 
(solid lines) $N_c$=4,8,12 (increasing value at $\omega = 0$) 
and DCA (dotted lines) $N_c$=4,8,12 (decreasing value at $\omega = 0$).}
\label{gwu4b5}
\end{figure}
\begin{figure}[b]
\includegraphics[width=\textwidth]{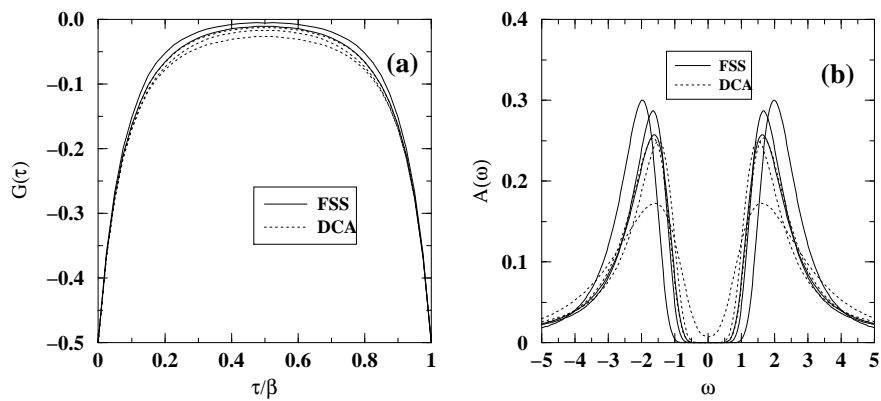}
\caption{  (a) Imaginary-time Green function at the Fermi point for 
$U=6$, $\beta=5$ for FSS
(solid lines) for $N_c$=4,8,16 (from top to bottom) and for DCA
 (dotted lines) for $N_c$=4,8,12 (from bottom to top). (b) The corresponding 
spectral weights for FSS
(solid lines) $N_c$=4,8,12 (broad to narrow gap) 
and DCA (dotted lines) $N_c$=4,8,12 (narrow to broad gap).  
Note that once a gap opens in the spectral function, 
the maximum-entropy analytic continuation procedure becomes 
unreliable.\cite{JARRELLandGUB}  This unreliability is believed to be the
source of the non-systematic nature of the peaks in the spectral functions.
Although unreliable for fine structure, the qualitative feature of the 
existence of a gap is accurately depicted in this figure.}
\label{gwu6b5}
\end{figure}

\begin{figure}[b]
\includegraphics[width=\textwidth]{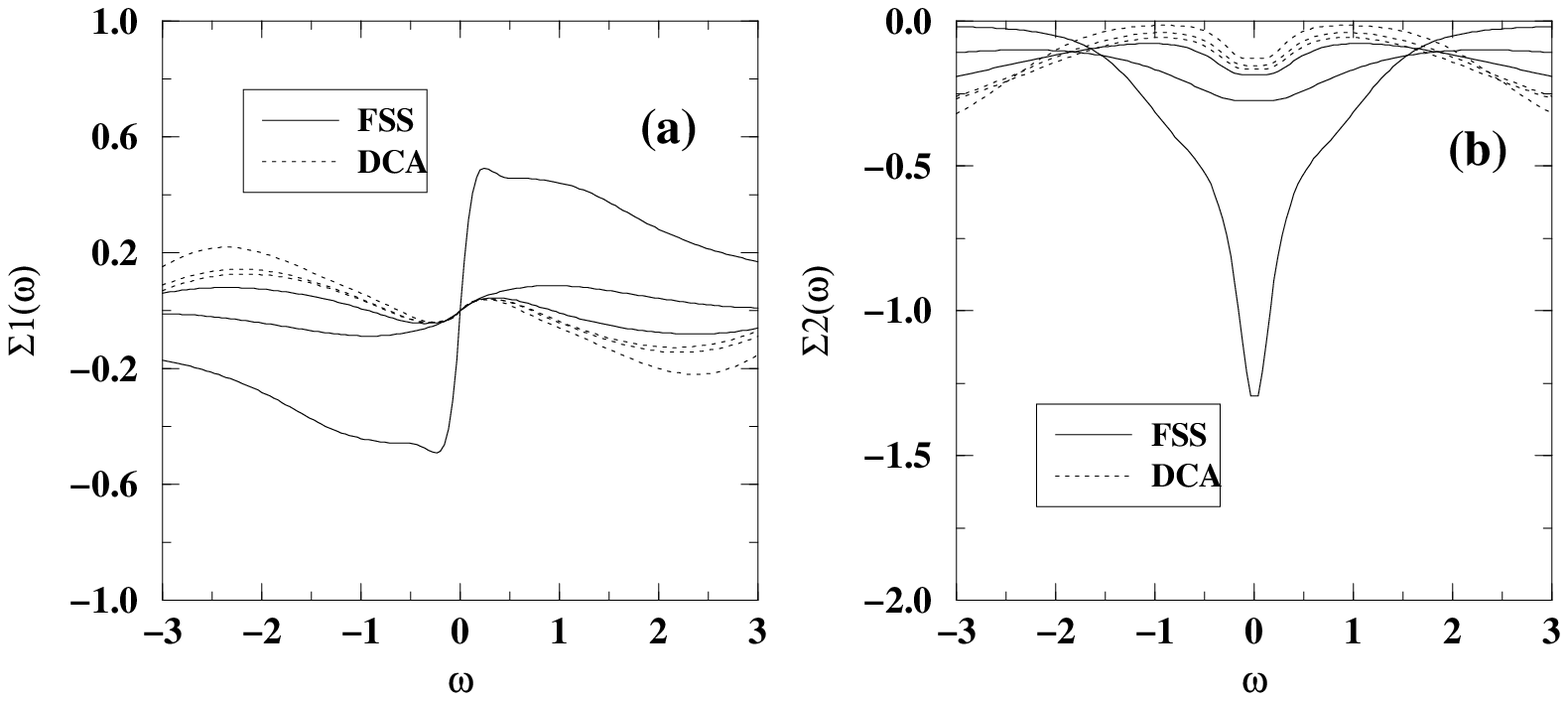}
\caption{  Real (a) and Imaginary (b) part of $\Sigma$ at the Fermi point 
for $U=2$, $\beta=5$ for FSS (solid lines)
for $N_c$=4,8,16 and for DCA (dotted lines). } 
\label{reimselfu2b5}
\end{figure}
\begin{figure}[b]
\includegraphics[width=\textwidth]{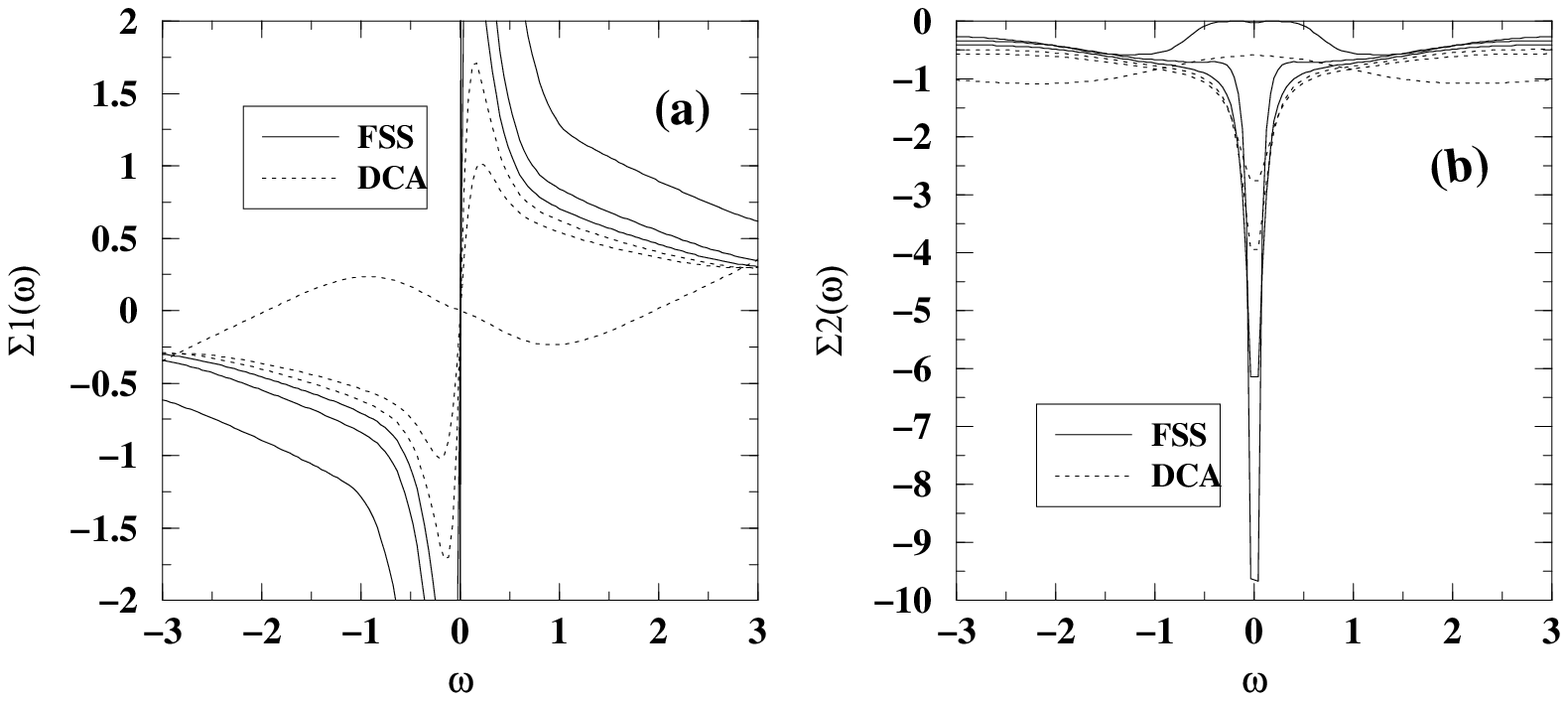}
\caption{  Real (a) and Imaginary (b) part of $\Sigma$ at the Fermi point 
for $U=4$, $\beta=5$ for FSS (solid lines)
for $N_c$=4,8,16 and for DCA (dotted lines). } 
\label{reimselfu4b5}
\end{figure}
\begin{figure}[b]
\includegraphics[width=\textwidth]{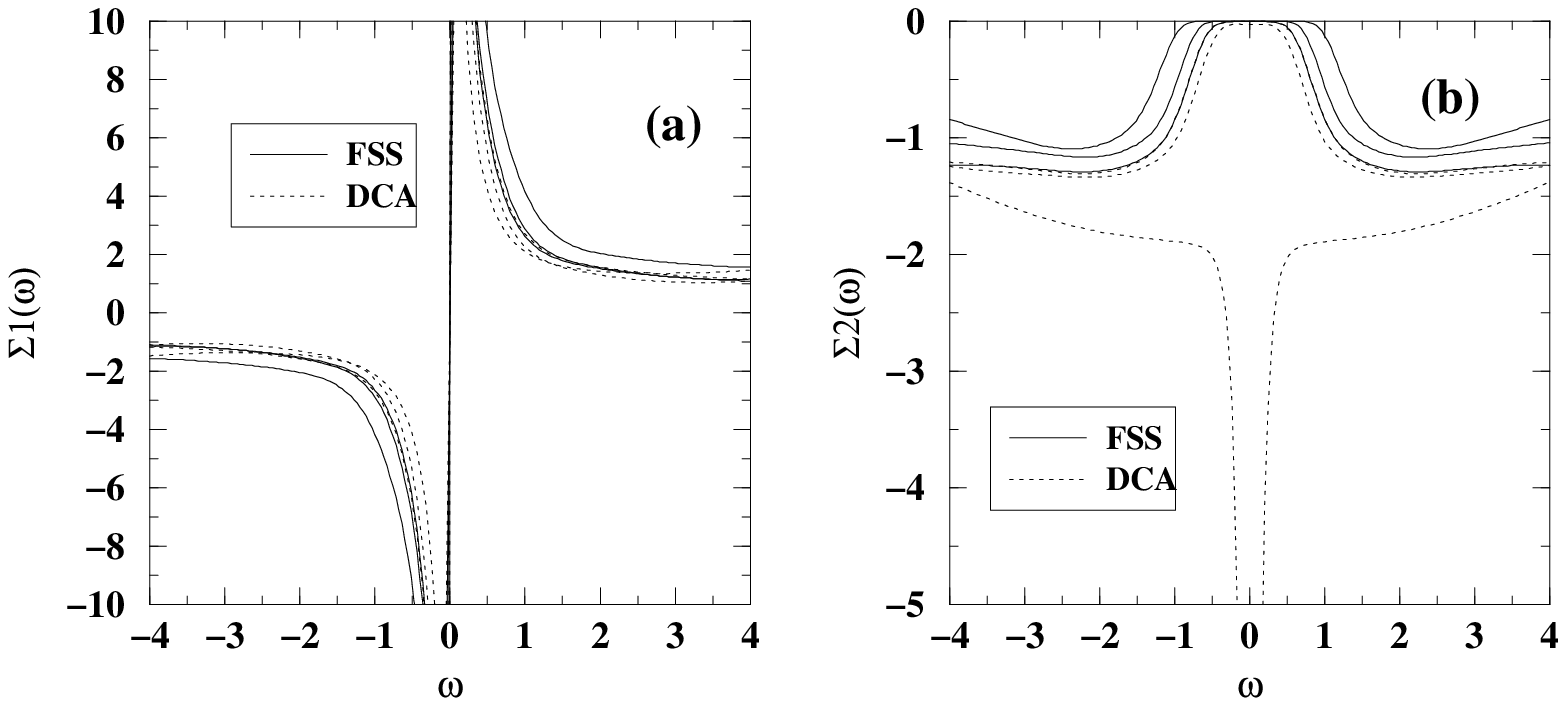}
\caption{  Real (a) and Imaginary (b) part of $\Sigma$ at the Fermi point 
for $U=6$, $\beta=5$ for FSS (solid lines)
for $N_c$=4,8,16 and for DCA (dotted line). } 
\label{reimselfu6b5}
\end{figure}
\section {Numerical Aspects}

One  difficulty encountered with the DCA algorithm is that a reliable
transform from imaginary-time quantities, in the QMC part, to Matsubara 
frequencies, for the coarse-graining part is needed.   A careful treatment 
of the frequency summation or the imaginary-time integration is crucial in 
order to ensure the accuracy and the stability of the algorithm and
to maintain the correct high-frequency behavior of the Green functions. 
We need to evaluate the following integral
\begin{eqnarray}
G_c({\bf K},i\omega_n) = \int_{0}^{\beta} d\tau e^{i\omega \tau}G_c({\bf K},\tau)  \quad\mbox{.} 
\end{eqnarray}
But from the QMC, we know the function $G_c({\bf K}, \tau)$ only at
a discrete subset of the interval $[0,\beta]$. As it may be readily seen
by discretizing the above equation, the estimation of $G_c({\bf K}, i\omega_n)$
becomes inaccurate at high-frequencies. This is formalized by Nyquist's theorem
which tells us that above the frequency $\omega_c= \frac{\pi}{\Delta \tau}$ 
unpredictable results are produced by conventional quadrature techniques. 
A straightforward way to cure this problem may be to increase the size of the 
set of $\tau$-points where the Green function is evaluated. But, this renders 
the QMC simulation rapidly intractable as seen in the previous section. A 
much more economic way to avoid the problem is to use an approximate method 
that is asymptotically exact.

Second-order perturbation theory is enough to obtain the correct asymptotic 
behavior. For instance, we compute the Matsubara-frequency Green function 
from the imaginary-time QMC Green function as follows \cite{jarrell}
\begin{eqnarray}
G_c({\bf K},i\omega_n)=  G_{c\,pt}({\bf K},i\omega_n) +
 \int_{0}^{\beta} d\tau e^{i\omega \tau}
(G_c({\bf K},\tau)-G_{c\,pt}({\bf K},\tau)) \quad\mbox{.} 
\end{eqnarray}
The integral is computed by first splining the difference
$G_c({\bf K},\tau)-G_{c\,pt}({\bf K},\tau)$, and then integrating the spline
(a technique often called oversampling).

Once convergence is reached, ${\bar{G}}=G_c$, and the QMC Green function 
$G_c$ may be analytically continued using the Maximum-entropy method (MEM).  
Unfortunately, there is no reliable way to perform the direct analytic 
continuation of $\Sigma({\bf K})$.  Pad\'e approximants lead to very 
unstable spectra because of the QMC statistical noise contained in 
$\Sigma(\bf K)$.  The binned imaginary-time Green function data accumulated
from the cluster calculation must be used to obtain lattice spectra
from which $\Sigma({\bf K})$ may be deduced.  To obtain the self-energy  
and spectral-weight function $A(\bf k, \omega)$ of the lattice in real 
frequencies, we first compute the cluster spectral-weight 
$\bar{A}(\bf K, \omega)$ by using the Maximum entropy
method \cite{JARRELLandGUB} for the inversion of the following integral equation
\begin{eqnarray}
\bar {G}({\bf K}, \tau) = \int d\omega \frac  {e^{-\omega \tau}}{1+
e^{-\beta \omega}} \bar {A}(\bf K, \omega)  \quad\mbox{,} 
\label{mem}
\end{eqnarray}
where $\bar {G}({\bf K}, \tau)$ is the 
imaginary-time Green function obtained
from the QMC simulation of the cluster. From  $\bar{A}({\bf K}, \omega)$, 
we obtain  the coarse-grained Green function in real frequencies
by Kramers-Kr\"onig analysis.
 Finally, we solve equation (1) to obtain $\Sigma({\bf K}, \omega)$ from
$\bar {G}({\bf K},\omega)$.

\section {Example: the One-dimensional Hubbard Model}

In this section, we apply the DCA to the one-dimensional Hubbard model 
at half-filling. Such a test is interesting for several reasons.  First, 
this test will help verify whether recent applications of the DCA in two 
dimensions are reasonably anticipated to be accurate.  In one dimension,
quantum fluctuations are stronger than in higher dimensions.  Hence, one 
intuitively expects the DCA to be less efficient in one than in higher 
dimensions.  So, if the DCA accurately captures the physics in one dimension, 
then it is highly likely to capture the physics of two and three dimensions 
accurately.  Second, although the DCA is known to become exact in the limit 
of an infinite cluster, an extensive, systematic analysis of the 
convergence of the DCA has not yet been performed.  Finally, it is desirable 
to illustrate differences between the DCA and a widely-applied FSS method in 
a well-studied, non-trivial problem.

The 1D Hubbard Hamiltonian is widely accepted as the most relevant model in 
the high temperature regime of the 1D organic materials, for which a 
significant volume of experimental data is available.  It is well-studied 
and provides a non-trivial test of the DCA.  The Hubbard Hamiltonian reads: 
\begin{eqnarray}
H=-t\sum_{i}(c_{i\sigma }^{+}c_{i+1\sigma }+hc)+U\sum_{i}n_{i\uparrow
}n_{i\downarrow }  \quad\mbox{,}  
\end{eqnarray}
with a next-nearest-neighbor hopping and an on-site repulsion $t$ and $U$, 
respectively, on a one-dimensional lattice. 
We set $t=1$ throughout this study and measure all energies in terms of $t$.
We work at half-filling, where the QMC is free of the fermion sign problem, 
eliminating one possible source of errors in both the FSS and the DCA.  This 
allows us to easily isolate actual discrepancies between the FSS and the DCA. 
Although the 1D Hubbard model can be solved exactly, the FSS QMC is currently
the only reliable method used to compute finite temperature dynamics. 
 
We now turn to a comparison of the imaginary-time Green function at the 
Fermi-point of the DCA and of FSS.  We see that the two methods converge 
differently to the exact result.  The quantity 
\begin{eqnarray}
G(k_{F},\beta/2)=\frac {1}{2}\int{  d\omega \frac{A(k_{F},\omega)}
{cosh(\frac{\beta \omega}{2})}}  \quad\mbox{,} 
\end{eqnarray}
is useful to measure the strength of correlations. At the Fermi-point
and at a fixed temperature $T$, it varies from $- \frac{1}{2}$ for $U=0$, 
an uncorrelated system, to $0$ for $U=\infty$, a highly-correlated system. 
In Fig. ~\ref{gwu2b5}(a), ~\ref{gwu4b5}(a), and ~\ref{gwu6b5}(a),
we display $G(\tau)$ for $U=2,4,6$; $N_c=4,8,16$; and $\beta=5$ for FSS and 
for the DCA.  These parameters were chosen to illustrate the three 
situations that are generic in the model: the metallic, the pseudogap, and 
the insulating regimes, respectively. In all cases, $G(k_{F},\beta/2)$ 
decreases with increasing cluster size for FSS while for the DCA it increases. 

This behavior marks the fundamental difference between FSS and the DCA.
At low temperatures, in FSS, the correlation length is greater than the 
lattice size. Thus, the effects of correlations are overestimated for 
smaller clusters because the systems are artificially closer to criticality 
than a system in the thermodynamic limit. This tendency is reduced by 
increasing the cluster size, which moves the system in the direction of the 
thermodynamic limit. The situation is radically different in the DCA where 
the system is already in the thermodynamic limit.  The DCA approximation 
restricts correlations to within the cluster length.  As the cluster size 
increases, longer range correlations are progressively included. Thus, the 
effects of the correlations increase with the size of the cluster.

This analysis is supported by the spectra shown in 
Fig.~\ref{gwu2b5}(b), ~\ref{gwu4b5}(b), and ~\ref{gwu6b5}(b).
For $U=2$, FSS shows a pseudogap for $N_{c}=4$. This
pseudogap vanishes at $N_{c}=8,16$ to become a peak at $\omega=0$. The
peak corresponding to $N_c=16$ is sharper that of $N_c=8$. In contrast,
 the DCA starts with a sharp peak at $N_c=4$. This peak progressively
broadens as $N_c$ is increased to $8$ and $16$. The FSS and the DCA peaks 
seem to converge to the same limit consistent with the results of $G(\tau)$.
For $U=4$, the system presents a pseudogap. The convergence to this
pseudogap is in conformity with the above analysis. The FSS system goes
from a gap to a pseudogap when $N_c$ is increased from $4$ to $16$. The
DCA evolves from a central peak at $N_c=4$ to a pseudogap at $N_c=16$. 
When $U=6$, the system is gapped; the FSS shows a large gap at 
$N_c=4$. This gap decreases for $N_c=8$ and $N_c=16$.

In Fig. ~\ref{reimselfu2b5}(a), ~\ref{reimselfu4b5}(a), 
and ~\ref{reimselfu6b5}(a),   
we show the real part of the self energy at $\beta=5$ for various 
interaction strengths, $U$.
For $U=2$ (Fig.~\ref{reimselfu2b5}), 
where a pseudogap exists for $N_c=4$, $\Sigma_1$ via FSS has
two solutions at non-zero $ \omega $ for the equation $\omega - \epsilon_k
-\Sigma_1(\omega)=0$. This happens when the slope  $\frac {dRe\Sigma(
\omega)}{d\omega}$ is greater than unity. This derivative decreases and
becomes smaller than 1 for $N_c=8,16$ corresponding to the single peak in
$A(\omega)$. In the DCA, $\frac {dRe\Sigma(\omega)}{d\omega}$ slowly increases
with $N_c$ but always remains smaller than 1.

The Imaginary part of the self energy $\Sigma_2(\omega)$ shown in Fig.
~\ref{reimselfu2b5}(b), ~\ref{reimselfu4b5}(b), and ~\ref{reimselfu6b5}(b),  
also has a monotonic behavior in which the limiting value is bracketed by
FSS and DCA results.  A general trend that emerges is that $\Sigma_2(\omega)$
has a local minimum in the vicinity of $\omega=0$ in the metallic
regime (Fig.~\ref{reimselfu2b5}), which minimum deepens sharply 
when a pseudogap appears (Fig.~\ref{reimselfu4b5}), and finally
$\Sigma_2(\omega=0)$ vanishes in the insulating regime 
(Fig.~\ref{reimselfu6b5}).

\section{Summary}
We have extensively analyzed the DCA by comparing it to FSS.  A coherent 
picture emerges from the investigation of the single-particle properties of 
the 1D Hubbard model.  By systematically underestimating the effects of
correlations, the DCA converges to the limiting value of a given quantity 
from a starting point that is opposite to the starting point of FSS.  Thus, 
the DCA can help resolve situations where one is unable to draw a conclusion
from FSS alone.  For example, in spite of considerable work with
FSS, it has not been possible to make a definitive statement as
to the existence of a pseudogap in the weak-coupling (U smaller than the
band-width) regime of the two-dimensional Hubbard model at half-filling.
It was impossible to get a definite answer from FSS
because, as expected, the pseudogap seemed to vanish when the
cluster size was increased \cite{white}. 
Recently, using DCA, Huscroft et al. showed that there is a pseudogap for
$4 \times 4$ to $8 \times 8$ clusters \cite{carey}.
From the analysis above,
since the DCA underestimates the effects of correlations, this
demonstrates unambiguously the existence of the pseudogap in the
half-filled Hubbard Model in the thermodynamic limit. 

The DCA has shown great promise and, even though a young method,
has already illustrated its great utility.  Still, much work remains
to be done.  One immediate application is the
the nature of the ground state of the two dimensional Hubbard
model; is it a Mott insulator or antiferromagnetic (Slater) insulator?
Another is the problem raised by Anderson concerning the
low energy behavior of 2D coupled doped Hubbard chains; are they 
Luttinger or Fermi liquids? \cite{anderson}  
Both of these problems have proven difficult to tackle using 
FSS approaches alone.
Finally, we have not discussed the convergence of the two-particle
quantities. 
A systematic comparison with FSS as in this work will be the subject 
of a future study.
 
\bigskip

\noindent{\bf Acknowledgments:} We wish to thank S. Allen for sharing his
BSS data.
 This work was supported by NSF grants DMR-9704021,
 DMR-9357199 and
by the Ohio Supercomputer Center. 
 

%

\end{document}